\documentclass[12pt]{iopart}
\usepackage{graphicx}
\usepackage{amsthm,amsfonts,amssymb}
\expandafter\let\csname equation*\endcsname\relax
\expandafter\let\csname endequation*\endcsname\relax
\usepackage{amsmath}
\usepackage{graphicx}
\usepackage{dcolumn}
\usepackage{latexsym}
\usepackage{epsfig}
\usepackage{bm}
\usepackage[all]{xy}

\begin{document}

\title[Gravity localization in a string-cigar braneworld]{Gravity localization in a string-cigar braneworld}
\author{Jose Euclides G. Silva, Victor Santos, and C. A. S. Almeida}

\address{Departamento de F\'{i}sica - Universidade Federal do Cear\'{a} \\ C.P. 6030, 60455-760
Fortaleza-Cear\'{a}-Brazil}

\begin{abstract}

We proposed a six dimensional string-like braneworld built from a warped product between
a 3-brane and the Hamilton cigar soliton space, the string-cigar braneworld. This transverse manifold is a well-known steady solution of the Ricci flow equation that describes the evolution of a manifold.
The resulting bulk is an interior and exterior metric for a thick string.
Furthermore, the source satisfies the dominant energy condition. It is possible to realize the geometric flow as a result of variations of the matter content of the brane, actually, as its tensions. Furthermore, the Ricci flow defines a family of string-like branes and we studied the effects that the evolution of the transverse space has on the geometric and physical quantities.
The geometric flow makes the cosmological constant and the relationship between the Planck masses evolves.
The gravitational massless mode remains trapped to the brane and the width of the mode depends on the evolution parameter.
For the Kaluza-Klein modes, the asymptotic spectrum of mass is the same as for the thin string-like brane and the analogue Schr\"{o}dinger potential also changes according to the flow.
\end{abstract}


\maketitle
\section{Introduction}
\indent \indent In the recent years it has been given much attention to the possibilities and consequences of the extra dimensions. In a five dimensional manifold, the Randall-Sundrum (RS) models
\cite{Randall:1999ee,Randall:1999vf} not only explained the hierarchy between the Planck mass $M_{P}$ and the Electroweak mass $M_{EW}$ but also predict a small correction to the Newtonian
potential. The cornerstone of these models is the warp product between the brane and the transverse manifold to the brane.

In six dimensions, these warped models are even more rich. Indeed, the two dimensional transverse space has its own geometry, curvature and symmetries. For instance, a
cylindrical symmetric space exterior to a string is conical with angular deficit proportional to the tension of the string
\cite{Olasagasti:2000gx,frolov,Geroch,Israel,Christensen,Gherghetta:2000qi}. In this matter many authors contributed
in recent years. Cohen and Kaplan proposed a warped solution between a flat brane and a cylinder produced by a global string \cite{Cohen:1999ia}. This solution has a
singularity far from the brane, what produces an effective compactification. Gregory \cite{Gregory:1999gv} has found a regular and stable solution allowing a time
dependence on the bulk geometry and adding a negative cosmological constant. Olasagasti and Vilenkin \cite{Olasagasti:2000gx} studied a large class of exterior solutions
produced by a global defect for all values of bulk and brane cosmological constant. On the other hand, Oda \cite{Oda:2000zc} has extended the
warped product between a $(p-1)$-brane and a hypersphere  whereas Carlos-Moreno proposed a warped solution which converges asymptotically to a cylinder, the so-called cigar-like universe \cite{deCarlos:2003nq}.

In Gherghetta-Shaposhnikov (GS) model \cite{Gherghetta:2000qi}, an infinite thin string-like brane is embedded in a six dimensional static and axisymmetric bulk.
In opposite to RS models, where the brane cosmological constant must be adjusted to the bulk cosmological constant, the hierarchy problem is solved using only the ratio
between the bulk cosmological constant and the string tension.
Moreover, the massless and massive graviton modes are localized in that thin brane and it is possible to resolve the boundary singularity by string methods and apply the AdS-CFT correspondence \cite{Ponton:2000gi}. However, as pointed out by Tinyakov and Zuleta \cite{Tinyakov:2001jt}, the thin limit
of the string like brane yields a source breaking the dominant energy condition.

Some authors proposed different solutions that do not possess this issue. Giovannini-Meyer-Shaposhnikov (GMS) proposed a string like braneworld builded from an Abelian vortex. They founded by numerical means a static, regular and complete (interior and exterior) solution satisfying the dominant energy condition \cite{Giovannini:2001hh}. Bostok \textit{et al} \cite{Bostock:2003cv} and Kofinas \cite{Kofinas:2004ae,Kofinas:2005py} proposed string-like defects with higher-order corrections. Kanno-Soda \cite{Kanno:2004nr}, Navaro \textit{et al} \cite{Navarro:2003vw,Navarro:2004di}, Papantonopoulos \cite{Papantonopoulos:2005ma,Papantonopoulos:2007fk}, Cline \textit{et al} \cite{Cline:2003ak}, and Vinet-Clinet \cite{Vinet:2004bk} proposed different thick string-like models and they studied mainly the cosmological consequences of these scenarios.

The models described above has two common features: the brane has an azimuthal symmetry about to the transverse space and the brane is generated by fields with fixed
physical parameters. In six dimensions, the bulk is builded from a warped product between the brane and a disc. Once the exterior geometry reflects the physical properties
of the string source, we argue that a geometric flow in the transverse space can be understood as a result of variation of physical parameters of the string.

In this spirit, we propose a complete, regular and static string-like solution. Instead of the disc, we have taken the warped product between the brane and Hamilton cigar
soliton, and we called the resulting solution as a string-cigar. The Hamilton cigar is a two dimensional regular manifold, conformal to the disc and asymptotically flat
\cite{chow}. This space is called a cigar due its cylindrical symmetry whose radius vanishes at the origin and it converges asymptotically to a cylinder
(cigar-shape) \cite{topping}. The name Hamilton soliton is due this manifold be a self-similar solution of the Ricci flow equation, first discovered by Richard Hamilton
\cite{hamilton3}. In another context, the Hamilton cigar soliton was also studied by Witten as a target space in String theory \cite{witten}.

The Ricci flow is a geometric flow of a manifold, where its metric evolves with some dimensionless parameter, according to a diffusion partial
differential equation \cite{chow,topping,hamilton3,hamilton1,hamilton2,MorganTian,Chow1,caochow,perelman,perelman2}. This equation is used to study the final manifold
that some initial manifold can reach evolving under this equation.
Ricci flow was first proposed by Hamilton \cite{hamilton1} in order to prove some the Poincar\'{e} conjecture \cite{MorganTian,perelman,perelman2}.

The Ricci flow has also applications in Physics \cite{Woolgar:2007vz}. In sigma models, the flow represents a first order approximation to the renormalization flow in the
target space \cite{Friedan:1980jf,Oliynyk:2005ak,Tseytlin:2006ak,Oliynyk:2007bv}. In Euclidean gravity, the Ricci flow can be used to prove the existence of Black hole
solutions \cite{Headrick:2006ti}. Moreover, since Arnowitt-Deser-Misner (ADM) mass is invariant under Ricci flow, this flux can be used to prove some statements about the
asymptotic behavior of the curved space-times \cite{xianzhe}. In three dimensional topological gravity, the Ricci flow is useful to study the decay of massive gravitons
\cite{Lashkari:2010iy}.

The choice of the Hamilton cigar as the transverse space yields an interesting extension of the GS model. On one hand, the string-cigar solution is defined inside and
outside the string defect. This enable us to study correction of GS model due the effects near the brane and inside the core of the string. For instance, this
thick string solution, unlike the GS model, satisfies the dominant energy condition. Besides, none of the string tensions is zero inside the core and the
relationships between them, that define the Tolman mass and the angular deficit, depend on the string width.

On the other hand, since the cigar soliton depends on an evolution parameter, we have a parameterized solution reflecting the changes of some physical parameters.
This allow us to study the effects that variations of the sources has on the geometric and physical quantities. In five dimensional braneworlds, parameter-dependent
domain walls are obtained by means of the deformed potentials \cite{Almeida:2009jc,Bazeia:2005hu}. In six dimensions, we proposed an exterior string solution evolving under a resolution flow, where the parameter comes from the $2$-cycle of the resolved conifold as the transverse space, a well-known smoothed orbifold of string theory \cite{Silva:2011yk,Candelas:1989js,p,Klebanov:2007us,Pando Zayas:2000sq,VazquezPoritz:2001zt}.

Furthermore, the massless gravitational mode is normalizable and the KK modes differ from the GS model only near the brane. Moreover, the relationship between the Planck masses also evolves upon the geometric flow. This could result in some interesting particle physics effects. In fact, the hierarchy between the weak and Planck scale would depends on the geometrical changes of the extra dimensions.

The task of find complete string solutions is hard since it is needed to solve a
system of coupled differential equations for the interior and exterior region and then matched them \cite{Christensen,Cohen:1999ia,frolov,Israel}.
The known solution have been performed numerically \cite{Giovannini:2001hh,Gregory:1999gv,deCarlos:2003nq}. Therefore, the use of a Ricci flow solution as a transverse space can also be an useful tool to obtain new solutions (analytically or numerically) and to study their stability.

The analysis of the effects on the braneworlds due non-standard transverse manifolds has already been addressed before in the literature. Indeed, Randjbar-Daemi and
Shaposhnikov has assumed the transverse
manifold as a Ricci-flat or an homogeneous space and they obtained trapped massless gravitational modes and chiral fermions as well \cite{RandjbarDaemi:2000ft}.
Kehagias proposed a conical tear-drop whose conical singularity drains the vacuum energy to the transverse space, explaining the small value of the cosmological constant
\cite{Kehagias:2004fb}. Further, Garriga and Porrati has shown the absence of self-tuning in compact six-dimensional braneworld taking a football-shape transverse space
\cite{Garriga:2004tq}. Gogberashvili \textit{et al} has achieved three-generation for fermions on a $3$-brane whose transverse space has an apple shape
\cite{Gogberashvili:2007gg}. In analysis of fermions, Duan \textit{et al} proposed the torus as a transverse manifold \cite{Duan:2006es}. Some authors also have studied
the field behavior for a transverse manifold as the smoothed versions of the conifold, the resolved \cite{VazquezPoritz:2001zt} and deformed
\cite{Brummer:2005sh, Firouzjahi:2005qs, Noguchi:2005ws}.

Another non-standard solution is the so-called cigar-like universe where the transverse space has a cylindrical shape whose radius shrinks as we move toward the brane \cite{deCarlos:2003nq}. The cigar-like universe interpolates between a interior solution (with conical behavior) and an exterior solution ($AdS_{5}\times S^{1}$). The string-cigar solution is also defined inside the core of the string-like brane and but it converges to a $AdS_{6}$ space far from the brane.

The paper is organized as follows. In section II, we have made a review of the definition, basic properties, important solution and physical application of the Ricci flow. Further, we defined the Hamilton cigar and sketched its basic properties. In section III, we builded the bulk geometry and we studied the main geometrical
properties. Furthermore, we analyze the Einstein equation, the corresponding stress-energy-momentum components and string tension and we studied how the hierarchy between mass
scales changes with $k$.
In section IV, we studied the behavior of the massless and massive gravitational modes upon this flow. In section V, some conclusions, final remarks and perspectives were outlined.


\section{The Ricci flow and the Hamilton cigar soliton}
\indent \indent In this section we shall discuss about the definition and the main geometrical and physical properties of the so-called Ricci flow and of one of its solution, the Hamilton cigar soliton.

Let $(M_{D},g_{\lambda})$ a D-dimensional manifold with a Riemannian metric $g_{\lambda}$, $\lambda \in \mathbb{R}$. For each $\lambda$, the manifold $M_{D}$ has a distinct local geometry. Suppose it could pass continuously from one configuration to another through the partial differential equation:

\begin{equation}
\label{ricciflow}
 \frac{\partial g_{ab}(\lambda)}{\partial \lambda}= -2 R_{ab}(\lambda).
\end{equation}
This geometric flow is called the Ricci flow due the Ricci tensor be the responsible for driven the ``evolution" of metric tensor.

The self similarity is a key property of Ricci flow. Indeed, consider a scale transformation on the metric tensor \cite{topping}

\begin{equation}
\tilde{g}(x,\lambda)=\xi g\left(x,\frac{\lambda}{\xi}\right).
\end{equation}
Since both $\frac{\partial\tilde{g}}{\partial\lambda}=\frac{\partial g}{\partial\lambda}$ and $\tilde{R}_{ij}=R_{ij}$ then the Ricci flow equation is unaltered through scale transformation. Scale invariance provides a relation between Ricci flow and RG flow \cite{Friedan:1980jf,MorganTian,perelman,topping,Woolgar:2007vz}.

Another important feature of the Ricci flow it that it tends to shrink manifolds of positive scalar curvature and stretches those of negative curvature. Indeed, for the sphere
\begin{equation}
R_{ij}=(n-1)g_{ij}.
\end{equation}
Under the Ricci flow, the metric evolves as \cite{chow,topping}
\begin{equation}
g(\lambda)=(1-2(n-1)\lambda)g_{0}(S^{n}).
\end{equation}
Since for $\lambda_{f}=\frac{1}{2(n-1)} \Rightarrow g(\lambda_{f})=0$, the sphere is contracted to a point in a finite time.
For the hyperbolic space $\mathbb{H}^{n}$, an Einstein space of negative curvature whose Ricci tensor is given by $R_{ij}=-(n-1)g_{ij}$, the space expands infinitely.

An extension of Einstein spaces are the so-called gradient Ricci solitons that satisfy the equation \cite{chow,topping,Chow1,perelman,perelman2}

\begin{equation}
\label{gradientriccisoliton}
R_{ij}+\nabla_{i}\nabla_{j}f=\zeta g_{ij},
\end{equation}
where the function $f$ is called the potential of Ricci flow.

There are three different kinds of gradient Ricci solitons. For $\zeta=0$ the solutions are called steady; if $\zeta<0$ we have an expanding Ricci soliton and for $\zeta>0$ a shrinking soliton. In this article we will focus on a special steady solution called the Hamilton cigar soliton \cite{chow,topping}.

The steady solitons have the remarkable property of being extremes of the Perelman energy. Indeed, the Perelman energy functional is defined as \cite{chow,topping,perelman,perelman2}

\begin{equation}
\mathcal{F}(g,f)=\int_{M}{(R+|\nabla f|^{2})e^{-f}\sqrt{g}d^{n}x}.
\end{equation}
Note that the Perelman energy functional is the Euclidean version of the low energy supergravity action for the gravitational and dilaton fields \cite{p,perelman}.

The Hamilton cigar $\mathcal{C}_{2}$ is a two dimensional steady solution of the Ricci flow equation (\ref{ricciflowequation}), where
\begin{equation}
\label{cigarmetric}
 ds^{2}_{\lambda}=\frac{1}{(e^{4\lambda}+r^{2})}(dr^{2} + r^{2} d\theta^{2}),
\end{equation}
and $r\in[0,\infty)$, $\theta\in [0,2\pi]$, $\lambda \in (-\infty,\infty)$ \cite{chow,topping,hamilton3,caochow}.

The metric $(\ref{cigarmetric})$ above defines a family of conformal metrics to the disc.
Using a new variable defined by $r=e^{2\lambda}\sinh{\rho}$, the metric (\ref{cigarmetric}) yields \cite{chow,caochow}

\begin{equation}
\label{cigarmetric2}
ds^{2}_{\lambda}=d\rho^{2}+\tanh^{2}{\rho}d\theta^{2}.
\end{equation}
The metric (\ref{cigarmetric2}) has been studied by Witten as a target space metric on sigma models \cite{witten}.

The scalar curvature of cigar soliton is

\begin{equation}
R = 4 \text{sech}^{2}{\rho}=4\frac{e^{4\lambda}}{r^{2}+e^{4\lambda}}.
\end{equation}
Therefore, that manifold has an everywhere non-negative scalar curvature and it converges to a cylinder asymptotically, and that is why it is called a cigar.

It is worthwhile to mention that the metric $(\ref{cigarmetric})$ is dimensionless. In order to leave it with its right dimension, $[g]=L^{2}$, we introduced the constants $R_{0}$ and $c$ , with dimension $[R_{0}]=[c]=L$, yielding to
\begin{equation}
\label{conformalmetric}
 ds^{2}_{\lambda}=\frac{R_{0}^{2}}{(c^{2}e^{4\lambda}+r^{2})}(dr^{2} + r^{2} d\theta^{2}).
\end{equation}

Now let us make the replacement
\begin{equation}
 ce^{2\lambda}=a.
\end{equation}
Since $\lambda \in [-\infty,\infty) \Rightarrow a\in [c,\infty)$ and hence $a$ has dimension $[a]=L$. As $c$ can be any non zero real number, the parameter $a$ is defined in the range $(0,\infty)$.
The parameter $a$ defines a family of two-dimensional manifolds and the variation of $a$ defines a flux from one to another.

In order to leave the metric $(\ref{conformalmetric})$ in a simpler form, let us make the change of variable
\begin{equation}
\label{changeofvariable}
 r=a \sinh{k\rho},
\end{equation}
where $k=\frac{1}{R_{0}}$. Since the flow have only one free parameter, we must relate $a$ with $k$. Thus, let us choose $k=\frac{1}{a} \Rightarrow R_{0}=a$.
Hereinafter, we shall call $k$ as the evolution parameter.

The change $(\ref{changeofvariable})$ leaves the metric $(\ref{conformalmetric})$ to the neatly form
\begin{equation}
\label{cigarsolitonmetric}
 ds^{2}_{k}=d\rho^{2}+\frac{1}{k^{2}}\tanh^{2}{k\rho}d\theta^{2}.
\end{equation}
Henceforward, we shall use $(\ref{cigarsolitonmetric})$ as the cigar soliton metric. 


\section{Bulk geometry}
\label{Bulk geometry}
Once described the geometry of the Hamilton cigar which we will use as a transverse space, let us build a six dimensional bulk $\mathcal{M}_{6}$ of the form $\mathcal{M}_{6}=\mathcal{M}_{4}\times \mathcal{C}_{2}$, where $\mathcal{C}_{2}$ is the
Hamilton cigar described in the last section and the $\mathcal{M}_{4}$ is a 3-brane embedded in $\mathcal{M}_{6}$.

The action for the gravitational field minimally coupled with some matter source is

\begin{equation}
\label{action}
  S_{g} =\int_{\mathcal{M}_{6}}{\left(\frac{1}{2K_{6}}R-\Lambda +\mathcal{L}_{m}\right)\sqrt{-g}d^{6}x},
\end{equation}
where $K_{6}=\frac{8\pi}{M_{6}^{4}}$ and $M_{6}^{4}$ is the six-dimensional bulk Planck mass. Note that in this convention, the bulk cosmological constant $\Lambda$ has dimension
$[\Lambda]=L^{-6}=M^{6}$.

Now, let us propose the following warped metric between the 3-brane and the Hamilton cigar, namely \cite{Gherghetta:2000qi,Giovannini:2001hh,Oda:2000zc,Olasagasti:2000gx,deCarlos:2003nq}

\begin{eqnarray}
\label{metricansatz}
ds^{2}_{6} & =  & \sigma(\rho,k)\hat{g}_{\mu\nu}(x^{\zeta})dx^{\mu}dx^{\nu}+d\rho^{2} + \gamma(\rho,k) d\theta^{2},
\end{eqnarray}
where

\begin{equation}
\label{warpfunction}
\sigma(\rho,k)=e^{-\left(k\rho - \tanh{k\rho}\right)}
\end{equation}
and,
\begin{equation}
\label{angularmetric}
\gamma(\rho,k)=\frac{1}{k^{2}}\left(\tanh{k\rho}\right)^{2}\sigma(\rho,k).
\end{equation}

\begin{figure}[htb] 
       \begin{minipage}[b]{0.48 \linewidth}
           \fbox{\includegraphics[width=\linewidth]{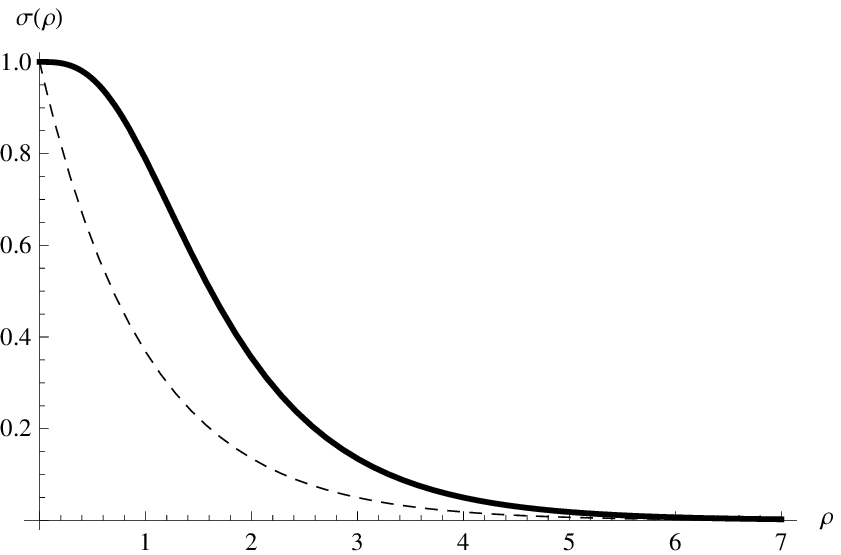}}\\
           \caption{Warp factor. For the string-cigar (thick line) and for the  and cigar-like universe (dashed line).}
           \label{variablechange}
       \end{minipage}\hfill
       \begin{minipage}[b]{0.48 \linewidth}
           \fbox{\includegraphics[width=\linewidth]{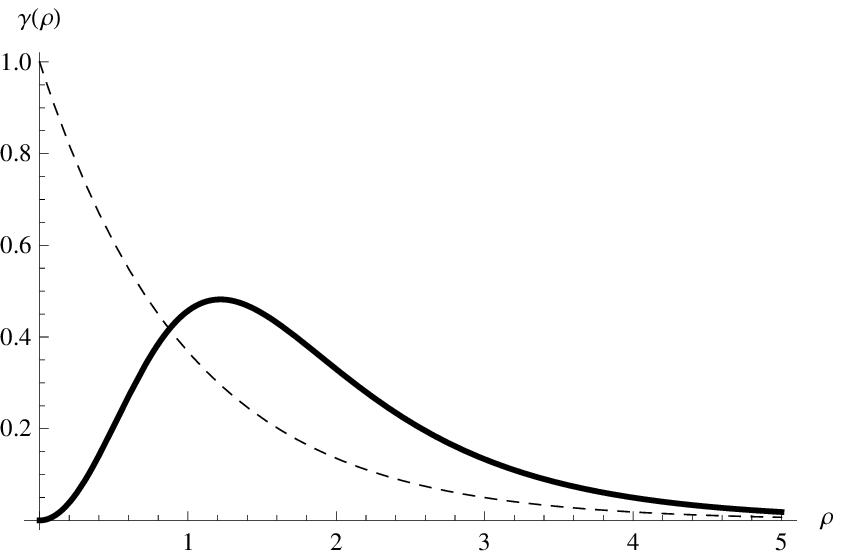}}\\
           \caption{Angular metric component. For exterior string defect (dashed line) and for the string-cigar model (thick line).}
           \label{changeofvariableinverse}
       \end{minipage}
   \end{figure}

This ansatz has a cylindrical symmetry about the 3-brane that lies in the point $r=0$. Furthermore, this metric represents a space-time inside and outside a
string-like defect. Indeed, the warp factors satisfy the usual string-like conditions for regularity at the origin \cite{Gherghetta:2000qi,Giovannini:2001hh,Tinyakov:2001jt,Israel}, namely
\begin{eqnarray}
 \sigma(0)=1 &, & \sigma'(0)= 0\\
 \gamma(0)=0 & , &(\sqrt{\gamma})'(0)=1,
\end{eqnarray}
where the prime $(')$ denotes the derivative according to $\rho$ variable. On the other hand, since $\lim_{\rho\rightarrow \infty}{\tanh \rho}=1$, that ansatz asymptotically goes to
the string-like exterior solution \cite{Olasagasti:2000gx,Gherghetta:2000qi,Gregory:1999gv,Oda:2000zc,deCarlos:2003nq,Tinyakov:2001jt,Giovannini:2001hh}.

The advantage of the metric ansatz proposed in $(\ref{warpfunction})$, and $(\ref{angularmetric})$ is that it provides geometric information for points inside
the core of the string defect, near and far from the brane. For core of the string we understand the region close to the string where the stress-energy-momentum tensor is more intense
while the far region is the vacuum. Since the stress-energy-momentum tensor decays slowly, there is an intermediary region, called near the string, where the stress-energy-momentum tensor
interpolate between those values.  As pointed out by many authors \cite{frolov,Geroch,Israel,Christensen}, the geometrical and physical properties are quite different on each region.

Close to the origin (in the core), the warp factor and the angular metric component has a $Z_{2}$ symmetry. It is worthwhile to say that the angular component has a conical behavior near the brane and decays exponentialy
far from the brane.

The warp factor has the same behavior of another axial complete solution called cigar-like universe \cite{deCarlos:2003nq}.
However, the angular metric components agree only in the core of the
string, where $\gamma(\rho)\approx \rho$. Far from the brane, the cigar-like geometry approaches to a cylinder whereas the string-cigar angular component tends to zero.

The scalar curvature for the metric ansatz (\ref{metricansatz}) is

\begin{eqnarray}
 R & = &
\frac{\hat{R}}{\sigma}-\Big[4\left(\frac{\sigma'}{\sigma}\right)' + 5\left(\frac{\sigma'}{\sigma}\right)^{2}+\left(\frac{\gamma'}{\gamma}\right)' + \frac{1}{2}
\left(\frac{\gamma'}{\gamma}\right)^{2}+2\frac{\sigma'}{\sigma}\frac{\gamma'}{\gamma} \Big]\nonumber\\
   & = & \frac{\hat{R}}{\sigma} - k^{2}\Big[\frac{15}{2}\tanh^{4}{k\rho}-16\tanh{k\rho}\text{sech}^{2}(k\rho)-4\text{sech}^{2}(k\rho)\Big].
\end{eqnarray}

\begin{figure}
\centering
\includegraphics[scale=1.1]{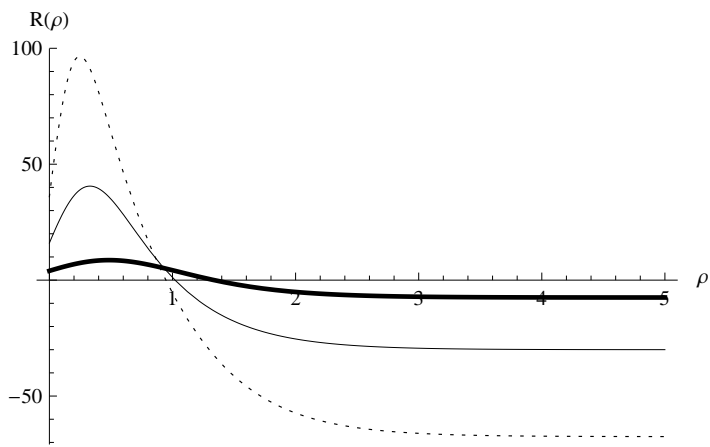}
\caption{Bulk scalar curvature. The manifold is smooth everywhere and it approaches to a $AdS_{6}$ asymptotically.}
\label{scalarcurvature}
 \end{figure}

The scalar curvature is sketched in the fig. (\ref{scalarcurvature}) for $\hat{R}=0$. Even though the angular metric factor shrinks the radius of the transverse circle the space-time has no conical singularity at the
origin. Inside the core of the string, the space-time has a positive scalar curvature that increases until reach a value and then decreases. Near the brane, the curvature turns to be negative and still decreasing.
In the far region, the scalar curvature reaches a constant and negative value what signals the space-times is asymptotically an $AdS_{6}$ manifold.
Besides, as more the value of parameter $k$ more the value of the curvature near and far from the brane. The dependence of the scalar curvature on the parameter $k$ is related with the k-dependence on the cosmological
constant, as we will see in the next section.

The geometric properties of the string-cigar solution is similar to the vortex solution found in \cite{Giovannini:2001hh}. Indeed, the metric components, scalar curvature and Ricci tensor have the same behavior.


\subsection{Einstein equations}

\indent \indent In this section we shall study some physical aspects of string-cigar geometry, e.g., the components of stress-energy-momentum tensor, the value of the cosmological constant, and the string tensions, through the Einstein
equation.

First of all, let us assume an axial symmetry ansatz for the stress-energy-momentum tensor \cite{Gherghetta:2000qi,Cohen:1999ia, Gregory:1999gv, Oda:2000zc, Tinyakov:2001jt, Giovannini:2001hh, Csaki:2000fc}
\begin{align}
\label{energymomentumansatz}
 T^{\mu}_{\nu} & = t_{0}(r)\delta^{\mu}_{\nu},\\
 T^{r}_{r} & = t_{r}(r),\\
 T^{\theta}_{\theta} & = t_{\theta}(r).
\end{align}
where,
\begin{equation}
T_{ab}=\frac{2}{\sqrt{-g}}\frac{\partial \mathcal{L}_{m}}{\partial g^{ab}}.
\end{equation}

Although we shall not propose a specific Lagrangian for the source, it is noteworthy to say that the ansatz of the stress-energy-momentum tensor above

\begin{equation}
 T_{ab}=T^{c}_{a}g_{cb}
\end{equation}
comprises the class of Lagrangians for fields minimally
coupled to gravity \cite{Csaki:2000fc}. Besides, the stress-energy-momentum form for a string-like defect satisfies this ansatz
\cite{Olasagasti:2000gx, Christensen, Gherghetta:2000qi, Cohen:1999ia, Gregory:1999gv, Tinyakov:2001jt, Giovannini:2001hh}.

The Einstein equations are
 \begin{equation}
\label{Einstein}
 R_{ab}-\frac{R}{2}g_{ab}=-K_{6}(\Lambda g_{ab}+T_{ab}),
\end{equation}
that for the metric ansatz in eq. (\ref{metricansatz}) yields the system of coupled differential equations \cite{Gherghetta:2000qi,Giovannini:2001hh,Tinyakov:2001jt}

\begin{eqnarray}
 \frac{3}{2}\left(\frac{\sigma'}{\sigma}\right)' + \frac{3}{2}\left(\frac{\sigma'}{\sigma}\right)^{2} + \frac{3}{4}\frac{\sigma'}{\sigma}\frac{\gamma'}{\gamma} + \frac{1}{4}\left(\frac{\gamma'}{\gamma}
\right)^ { 2 } +\frac {1} {2} \left(\frac { \gamma'}{\gamma}\right)'\\ \nonumber =-K_{6}(\Lambda+t_{0}(\rho))+\frac{K_{4}\Lambda_{4}}{\sigma},\\
\frac{3}{2}\left(\frac{\sigma'}{\sigma}\right)^{2}+\frac{\sigma'}{\sigma}\frac{\gamma'}{\gamma}=-K_{6}(\Lambda+t_{\rho}(\rho))+\frac{2K_{4}
\Lambda_ { 4 } } { \sigma }, \\
2\left(\frac{\sigma'}{\sigma}\right)'  +  \frac{5}{2}\left(\frac{\sigma'}{\sigma}\right)^{2}= -K_{6}(\Lambda+t_{\theta}(\rho))+\frac{2K_{4}\Lambda_{4}}{\sigma},
\end{eqnarray}
where, we have used the prime $(')$ to represent the derivative $\frac{d}{d\rho}$. Furthermore, the four-cosmological constant on the 3-brane satisfies
\begin{equation}
 \hat{R}_{\mu\nu}-\frac{\hat{R}\hat{g_{\mu\nu}}}{2}=-K_{4}\Lambda_{4}\hat{g}_{\mu\nu},
\end{equation}
where, $K_{4}=\frac{8\pi}{M_{4}^{2}}$. However, we have chosen a metric ansatz where $\lim_{r\rightarrow\infty}{\sigma(\rho)=0}$, and then the Einstein equations will blow up at infinity. Some authors solved this question by allowing a width to the brane and modifying the warp functions \cite{Kanno:2004nr,Navarro:2003vw,Navarro:2004di,Papantonopoulos:2005ma,Papantonopoulos:2007fk,Vinet:2004bk} or adding higher-order terms \cite{Bostock:2003cv,Kofinas:2004ae,Kofinas:2005py}. Here, following the flat-brane configurations \cite{Gherghetta:2000qi,Giovannini:2001hh,Tinyakov:2001jt,Ponton:2000gi}, henceforward we shall set $\Lambda_{4}=0$.

\begin{figure}
\centering
\includegraphics[scale=1.1]{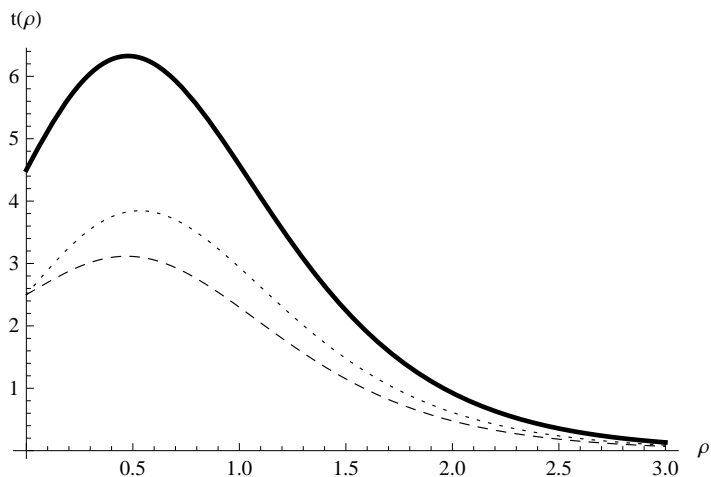}
\caption{Components of energy-momentum tensor. The density of energy (thick line) is always greater than the angular pressure (dotted line) and radial pressure (dashed
line).}
\label{plotenergymomentum}
 \end{figure}

Using the metric ansatz (\ref{metricansatz}), we have found the components of the stress-energy-momentum tensor, namely
\begin{eqnarray}
\label{energymomentumtensor}
t_{0}(\rho,k) & = & \frac{k^{2}}{K_{6}}\Big( 7\text{sech}^{2}{k\rho}+\frac{13}{2}\text{sech}^{2}{k\rho}\tanh{k\rho}-\frac{5}{2}\text{sech}^{4}{k\rho}\Big)\\
 t_{\rho}(\rho,k) & = & \frac{k^{2}}{K_{6}}\Big( 5\text{sech}^{2}{k\rho}+ 2\text{sech}^{2}{k\rho}\tanh{k\rho} - \frac{5}{2}\text{sech}^{4}{k\rho}\Big)\\
 t_{\theta}(\rho,k) & = & \frac{k^{2}}{K_{6}}\Big(5\text{sech}^{2}{k\rho}+4\text{sech}^{2}{k\rho}\tanh{k\rho}-\frac{5}{2}\text{sech}^{4}{k\rho}\Big).
\end{eqnarray}
These functions was plotted in figure (\ref{plotenergymomentum}). It is worthwhile to say that all the components have compact support near the origin where the 3-brane is placed. Furthermore, the components satisfy the
dominant, strong and weak energy condition that turn this geometry an extension of the GS model.

Besides, from the graphic we can define the core, near and far zones. Indeed, for $\rho > 3$ the components of energy-tensor are almost zero. Then we can define this region as the vacuum (far zone).
For $\rho \approx 1.5$ all the components reach half of their maximum value and then, for $0\leq \rho \leq 1.5$ we define a zone as the core of the string defect with width $\epsilon \approx 1.5$.
The intermediary region is the so called near zone. Note that for $\rho \approx 1.5$ the angular metric component, the scalar curvature
has a change of its behavior.

It is worth mention the behavior of the energy density $t_{0}$ in the core of the string-cigar solution is similar to one found in \cite{Giovannini:2001hh} for a vortex solution with winding number $n=2$.
Indeed, this component grows from the origin reaching a maximum and then it decays. On the other hand the radial and angular components, unlike the solution in \cite{Giovannini:2001hh}, behave as the energy density.

The difference between the string-cigar solution and the vortex model \cite{Giovannini:2001hh} can have many origins, among them: the string-cigar can be a solution for an Abelian
Einstein-Maxwell-Higgs model with higher winding number or a different symmetry breaking potential; further, we argue that as in cigar-universe solution, it can be a solution for a supersymmetric theory
\cite{deCarlos:2003nq}.

Since the components of energy momentum tensor depend on the $k$, the maximum value of these functions also depends on the evolution parameter. Then, the geometric flow alters the width of the string core $\epsilon$.

Another consequence of this geometry is that the bulk has a negative and parameter dependent cosmological constant. Indeed, constant $k$ is related with the cosmological constant of the bulk by the well-known relation \cite{Gherghetta:2000qi,Oda:2000zc}
\begin{equation}
 k^{2}=-\frac{2K_{6}}{5}\Lambda \Rightarrow \Lambda=\Lambda(k).
\end{equation}
Therefore, the geometric flow represents a variation of the bulk cosmological constant. This is an expected result since the asymptotic value of scalar curvature depends on the evolution parameter $k$ and $R=3\Lambda$.


\subsection{String tensions}

\indent \indent In this subsection we studied the effects of the geometric flow has on the string tensions. We also have shown that these quantities depend on the string width.

We defined the 4-tension per unit of volume of the 3-brane as \cite{Gherghetta:2000qi,Oda:2000zc,Tinyakov:2001jt,Giovannini:2001hh}

\begin{eqnarray}
 \mu_{i}(k) & = & \int_{0}^{\infty}{t_{i}(\rho,k)\sigma^{2}(\rho,k)\sqrt{\gamma(\rho,k)}d\rho}\nonumber\\
            & = & \frac{1}{k}\int_{0}^{\infty}{t_{i}(\rho,k)e^{-\frac{5}{2}(k\rho-\tanh{k\rho})}\tanh{k\rho}d\rho}.
\end{eqnarray}
Note the string tensions are all finite, smooth and have compact support around the brane, as expected. Moreover, the higher the value of the $k$ more located are the tensions. This indicates the string width $\epsilon$ varies with $k$.

It is worth mention here that the geometric flow is related to a variation of physical quantities, as the string tensions. Therefore, let us see how the string-cigar solution alters the Tolman mass and the string angular deficit computed in the GS model. Once the stress-energy-momentum tensor vanishes smoothly, we shall consider the integration range $[0,\epsilon]$, where $\epsilon$ is the width of the core. From Einstein equations we obtain

\begin{equation}
\label{tolmanmass}
(\sigma\sigma'\sqrt{\gamma})|_{0}^{\epsilon}=-\frac{1}{M_{6}^{4}}\left(\frac{1}{2}(\mu_{\rho}+\mu_{\theta})+ \frac{\Lambda}{2}\int_{0}^{\epsilon}{\sigma^{2}\sqrt{\gamma}d\rho} -4\Lambda_{4}\int_{0}^{\epsilon}{\sigma\sqrt{\gamma}d\rho} \right),
\end{equation}
and
\begin{equation}
\label{angulardeficit}
\sigma^{2}(\sqrt{\gamma})'|_{0}^{\epsilon} = -\frac{1}{M_{6}^{4}}\left( (\mu_{0}+\frac{1}{4}\mu_{\rho}-\frac{3}{4}\mu_{\theta})+\frac{\Lambda}{2}\int_{0}^{\epsilon}{\sigma^{2}\sqrt{\gamma}d\rho} -\Lambda_{4}\int_{0}^{\epsilon}{\sigma\sqrt{\gamma}d\rho} \right).
\end{equation}

The term $\mu_{\rho}+\mu_{\theta}$ is known in the literature as the Tolman mass of the string \cite{frolov,Gherghetta:2000qi,Tinyakov:2001jt}. Thus, eq. (\ref{tolmanmass}) tell us the Tolman mass evolves with the evolution parameter $k$ and it depends on the bulk and string cosmological constant. Since eq. (\ref{angulardeficit}) depends on the derivative of the angular metric component, it provides a measure of the angular deficit of the exterior space-time surrounding the string \cite{frolov,Gherghetta:2000qi,Tinyakov:2001jt}. Then, the angular deficit also depends on the cosmological constant terms.

From eqs. (\ref{tolmanmass}) and (\ref{angulardeficit}) we obtain the relation
\begin{equation}
\label{tensionrelation}
\mu_{0}-\mu_{\theta}=M_{6}^{4}\tanh^{3}{(k\epsilon)}\sigma^{\frac{5}{2}}(\epsilon).
\end{equation}
The relationship (\ref{tensionrelation}) is analogous to that in GS model \cite{Gherghetta:2000qi}. Nevertheless, in string-cigar solution, the relation depends on the cosmological constant ($k$) and on the string width $\epsilon$.

Those consequences are due the nonzero value of the core width and they are valid even for the flat-brane $\Lambda_{4}=0$. Indeed, for the GS model, i.e., for the thin string limit $\epsilon \rightarrow 0$, the terms of cosmological constant drop out of the equations (\ref{tolmanmass}) and (\ref{angulardeficit}). In this limit we do not need tune the brane cosmological constant to the bulk cosmological constant, as in RS models \cite{Randall:1999ee,Randall:1999vf}. Although, as shown in \cite{Tinyakov:2001jt}, the thin string limit contradicts the dominant energy condition.


\subsection{Mass hierarchy}

In the last subsection, we have seen that string tensions, including the string mass, depend on the evolution parameter $k$. In this section we have analyzed how the geometric evolution alters the relation between the bulk and brane mass scale.

In this geometry, the relationship between the four-dimensional Planck mass ($M_{4}$) and the bulk Planck mass
($M_{6}$) is given by

\begin{eqnarray}
M^{2}_{4} & = & 2\pi M_{6}^{4}\int_{0}^{\infty}{\sigma(\rho,k)\sqrt{\gamma(\rho,k)}d\rho}\nonumber\\
          & = & \frac{2\pi M_{6}^{4}}{k}\int_{0}^{\infty}{e^{-\frac{3}{2}(k\rho-\tanh{k\rho})}\tanh{k\rho}d\rho}.
\label{planckmass}
\end{eqnarray}

Since all the metric components are limited functions,
\begin{equation}
\int_{0}^{\infty}{e^{-\frac{3}{2}(k\rho-\tanh{k\rho})}\tanh{k\rho}d\rho} \approx \frac{1}{k}.
\end{equation}
Then, $M_{4}^{2}\approx 2\pi M_{6}^{4}\frac{1}{k^{2}}$. Hence, this geometry can be used to tune the ratio between the Planck masses, explaining the hierarchy between them.

Note, however, that the relationship between the Planck masses depends on the evolution parameter $k$. Hence, a evolution of the bulk geometry could alter the hierarchy between the fundamental forces. For large value of $k$, the brane Planck mass is smaller than the bulk Planck mass. In order to obtain $M_{4} \gg M_{6}$, we must do $k\rightarrow 0$. This is an extension of the GS tuning of the Planck masses \cite{Gherghetta:2000qi,Oda:2000zc} for points inside and near the core. Indeed, for points outside the string core, i.e., for $\rho \rightarrow \infty$, the hyperbolic function $\tanh{k\rho}\approx 1$, and we obtain the GS mass tuning \cite{Gherghetta:2000qi}.


\section{Gravity localization}
\label{Gravity localization}

\indent \indent Now let us study the localization of small perturbations of the background metric around a flat brane in the geometry analyzed so far. The perturbation is such that
\begin{equation}
ds^{2}_{6}=\sigma(\rho,k)(\eta_{\mu\nu}+h_{\mu\nu}(x^{\zeta},\rho,\theta,k))dx^{\mu}dx^{\nu}+d\rho^{2}+\gamma(\rho,k) d\theta^{2}.
\end{equation}

Using the traceless transverse gauge

\begin{equation}
 h^{\mu}_{\nu}=\nabla_{\mu}h^{\mu\nu}=0,
\end{equation}
the linearization of the Einstein equations (\ref{Einstein}) with the source given by eq.(\ref{energymomentumansatz}) yields a decoupled equation for the gravitational perturbation, namely \cite{Cohen:1999ia, Gherghetta:2000qi,Giovannini:2001hh, Csaki:2000fc}
\begin{equation}
\label{pertubationequation}
\Box_{6} h_{\mu\nu}=\partial_{a}(\sqrt{-g_{6}}g^{ab}\partial_{b}h_{\mu\nu})=0.
\end{equation}


Let us assume that the symmetric tensorial field $h_{\mu\nu}$ is a product of a 4-component field with Poincar\'{e} symmetry on the 3-brane $\hat{h}_{\mu\nu}$ and another scalar field living only in the transverse space,
the well-known process called Kaluza-Klein (KK) decomposition \cite{Gherghetta:2000qi,Cohen:1999ia,Gregory:1999gv,Oda:2000zc,Giovannini:2001hh}

\begin{equation}
 h_{\mu\nu}(x^{\zeta},\rho,\theta)=\hat{h}_{\mu\nu}(x^{\zeta})\tilde{\phi}(\rho,\theta).
\end{equation}

From Poincar\'{e} symmetry the tensorial field on the 3-brane $\hat{h}_{\mu\nu}$ must satisfies the mass condition
\begin{equation}
 \Box_{4}\hat{h}_{\mu\nu}(x^{\zeta}) = - m^{2}\hat{h}_{\mu\nu}(x^{\zeta}).
\end{equation}
Since $0\leq \theta \leq 2\pi$, let us assume that $\tilde{\phi}(\rho,\theta)$ can be expanded in Fourier series as

\begin{equation}
\label{transverseansatz}
 \tilde{\phi}(\rho,\theta)=\chi(\rho)\sum_{l=0}^{\infty}{e^{il\theta}}.
\end{equation}
Using the ansatz $(\ref{transverseansatz})$, eq. (\ref{pertubationequation}) yields the differential equation for the transverse component

\begin{equation}
\label{stequation1}
 \left(\sigma^{\frac{5}{2}}\sqrt{\beta}\chi'(r)\right)'+\sigma^{\frac{3}{2}}\sqrt{\beta}\left(m^{2}-\frac{l^{2}}{\beta}\right)\chi(\rho)=0,
\end{equation}
where $\beta(\rho,k)=\frac{\tanh^{2}{k\rho}}{k^{2}}$.

Equation (\ref{stequation1}) is a Sturm-Liouville like equation. Further, let us looking for solutions that satisfy the boundary conditions \cite{Gherghetta:2000qi}
\begin{equation}
\label{boundaryconditions}
\chi'(0)= \lim_{\rho\rightarrow\infty}\chi'(\rho)=0.
\end{equation}

Given two solutions of eq. (\ref{stequation1}), namely $\chi_{i}(\rho)$ and $\chi_{j}(\rho)$, the orthogonality relation between them is given by

\begin{equation}
\label{ortogonality}
 \int_{0}^{\infty}{\sigma(\rho,a)^{\frac{3}{2}}\sqrt{\beta(\rho,k)}\chi_{i}^{\ast}\chi_{j}d\rho}=\delta_{ij}.
\end{equation}

We can rewrite eq. (\ref{stequation1}) as
\begin{equation}
\label{radialequation}
 \chi''(\rho)+\left(\frac{5}{2}\frac{\sigma'}{\sigma}+\frac{1}{2}\frac{\beta'}{\beta}\right)\chi'(\rho)+
 \frac{1}{\sigma}\left(m^{2}-\frac{l^{2}}{\beta}\right)\chi(\rho)=0.
\end{equation}
Note that equation (\ref{radialequation}) is similar to that
found in string-like geometries \cite{Gherghetta:2000qi,Oda:2000zc}, regardless the cigar term $\beta(\rho,k)$.
Besides, we could define an effective angular
number $l_{eff}=\frac{l^{2}}{\beta(\rho,k)}$ which would depends on the distance from the brane and on the resolution parameter.


\subsection{Massless mode}
For $m=0$, a constant function is a solution of eq. (\ref{radialequation}). Thus, from orthogonality relation (\ref{ortogonality}), we can define the zero-mode as
\begin{equation}
\chi_{0}(\rho,k)=N\sigma(\rho,k)^{\frac{3}{4}}\left(\frac{\tanh{k\rho}}{k}\right)^{\frac{1}{4}},
\end{equation}
where
\begin{equation}
N^{2}=\int_{0}^{\infty}{\sigma(\rho,k)^{\frac{3}{2}}\left(\frac{\tanh{k\rho}}{k}\right)^{\frac{1}{2}}d\rho}.
\end{equation}

\begin{figure}
\centering
\includegraphics[scale=0.5]{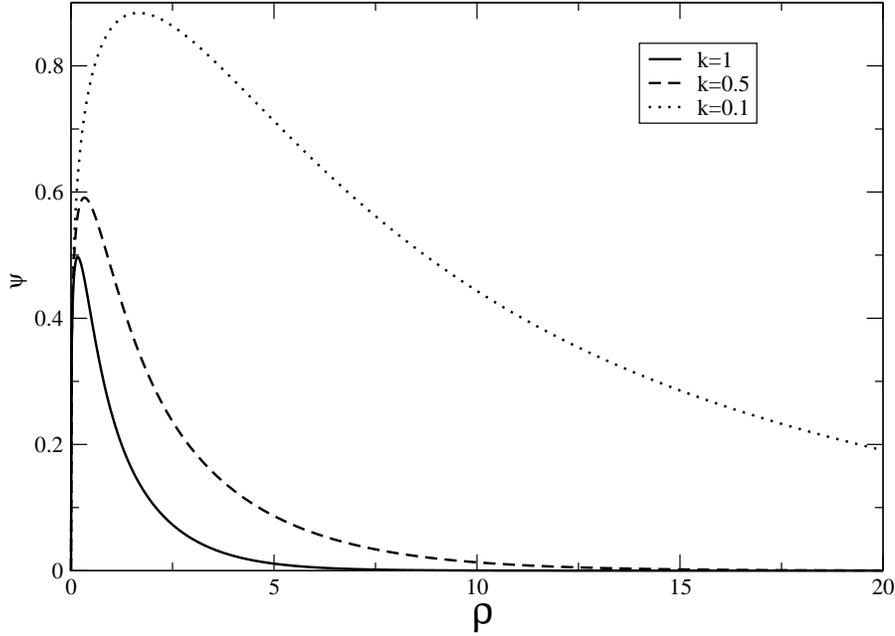}
\caption{Gravitational massless mode. This zero-mode is trapped to the brane for any value of $k$. Indeed, for $r>3$, the eigenfunction has an exponential decreasing behavior. In addition, the higher the value of $k$ higher the rate the massless mode vanishes. Besides, at the interior of the string, the lower the value of $k$ the higher the amplitude of the eigenfunction.}
\label{Massless mode.}
 \end{figure}

Since the massless mode has compact support, as can be seen in the figure (\ref{Massless mode.}), it is normalizable and then, we claim that the gravitational field is trapped around the string-like 3-brane.


\subsection{Massive modes}

\indent \indent Now let us study the KK modes of eq. (\ref{radialequation}). Using the expressions for the metric factor yields
\begin{equation}
\label{massivemodeequation2}
\chi''+\Big[-\frac{5k}{2}+k\text{sech}^{2}(k\rho)\left(\frac{5}{2}+\frac{2}{\tanh{k\rho}}\right)\Big]\chi'+e^{(k\rho-\tanh{k\rho})}\left(m^{2}-\frac{l^{2}k^{2}}{
\tanh^ { 2 } { k\rho}} \right)\chi=0.
\end{equation}

Eq.$(\ref{massivemodeequation2})$ together with the boundary conditions $(\ref{boundaryconditions})$ is a complex problem to attack directly and analytically. Hence, we shall concern ourselves to study
this equation in the near and far from the brane regimes.

In the limit $\rho \rightarrow \infty$ (vacuum), eq. (\ref{massivemodeequation2}) turns to be

\begin{equation}
\label{gherghettaequation}
\chi''-\frac{5k}{2}\chi'+e^{(k\rho)-1}\left(m^{2}-l^{2} \right)\chi=0.
\end{equation}
Equation (\ref{gherghettaequation}) is analogous to the massive modes equation of GS model \cite{Gherghetta:2000qi} with a mass term rescaled as
$m \rightarrow e^{\frac{-1}{2}}m$. The solution of eq. (\ref{gherghettaequation}) can be written in terms of the Bessel function as

\begin{equation}
\chi(\rho)=e^{\frac{5k\rho}{4}}\Big[C_{1}J_{\frac{5}{2}}\left(\frac{2m'}{k}e^{\frac{k\rho}{2}}\right)+C_{2}Y_{\frac{5}{2}}\left(\frac{2m'}{k}e^{\frac{k\rho}{2}}\right)\Big].
\end{equation}
where $´m'=\frac{m}{\sqrt{e}}$. Therefore, we argue that, for asymptotic points, the Kaluza-Klein spectrum is similar to the GS model \cite{Gherghetta:2000qi}.

In the limit $\rho\rightarrow 0$, i.e., for points close to the brane, let us make the change of variable $x:[0,\infty)\rightarrow [0,1]$ given by
\begin{eqnarray}
\label{xchange}
 x & = & \tanh{k\rho}\\
 y(x) & = & \chi(\rho)
\end{eqnarray}
That transformation allows that eq. (\ref{massivemodeequation2}) can be expressed as

\begin{eqnarray}
\label{massivemodenear}
 y''(x)+\Big[\frac{4x^{2}-5x^{3}-8x^{2}+4}{2x(1-x^{2})^{2}}\Big]y'(x) & = & -\frac{e^{(-x+\tanh^{-1}{x})}}{(1-x^{2})^{2}}\mu^{2}y(x).
\end{eqnarray}

Let us looking for the solution of eq. (\ref{massivemodenear}) in a Taylor series form

\begin{equation}
 y(x) = y(0) + y^\prime(0)x + \frac{y^{\prime\prime}(0)}{2}x^2
  + \frac{y^{(3)}(0)}{3!}x^3+\frac{y^{(4)}(0)}{4!}x^4 +
  \frac{y^{(5)}(0)}{5!}x^5+\mathcal{O}(x^6).
\end{equation}
From the boundary conditions expressed in eq. (\ref{boundaryconditions}), we obtain $y'(0)=0$. Substituting the resulting series on the differential equation (\ref{massivemodenear}) and retaining only terms up to $\mathcal{O}(x^{4})$ we obtain the approximated solution

\begin{equation}
y(x)=y(0)\left(1-\frac{\mu^{2}}{6}x^{2}+\frac{\mu^{2}(\mu^{2}-12)}{120}x^{4}-\frac{7\mu^{2}}{180}x^{5}\right).
\end{equation}
Hence, the massive mode is smooth in the core and near the brane.


\subsection{Analogue quantum potential}
Another way to study the massive modes lies on turn eq. (\ref{radialequation}) into a Schr\"{o}dinger-like equation and study its analogue quantum potential \cite{Cohen:1999ia}.

\begin{figure}
\centering
\includegraphics[scale=0.5]{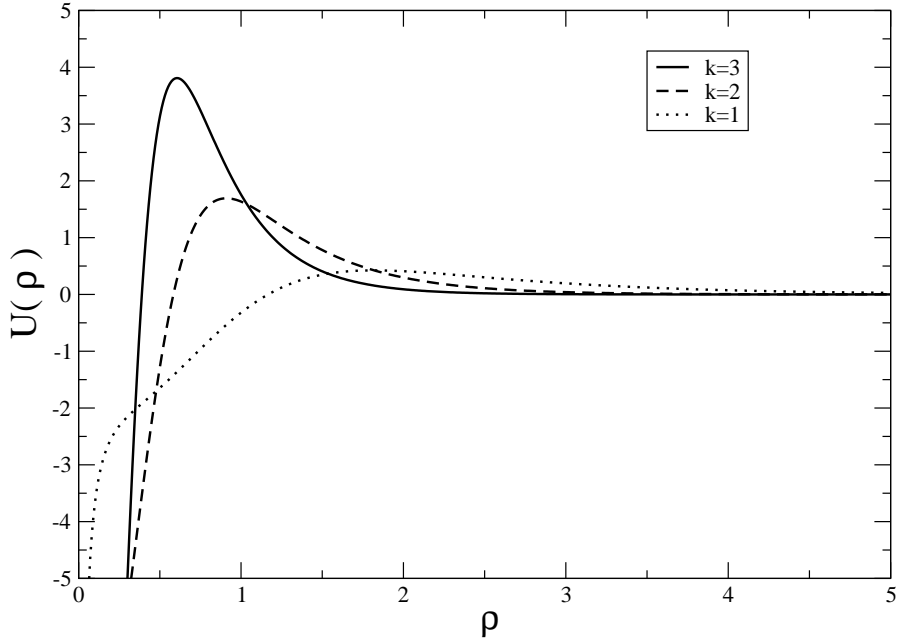}
\caption{Potential for the linear warp factor. The graphic has a well-known shape called \textit{volcano potential} for any $k$. The higher the value of $k$ more high is the barrier.}
\label{cigarpotential}
 \end{figure}

Firstly, let us perform the change of variable $z=z(\rho)$, namely
\begin{equation}
\label{change2}
z=z(\rho)=\int^{\rho}{\sigma^{-\frac{1}{2}}}d\rho'.
\end{equation}
Furthermore, let us write $\chi(z)$ in the form
\begin{equation}
 \chi(z)=u(z)\Psi(z).
\end{equation}
Making
\begin{equation}
\frac{\dot{u}}{u}=-\frac{1}{2}\left(2\frac{\dot{\sigma}}{\sigma}+\frac{1}{2}\frac{\dot{\beta}}{\beta}\right),
\end{equation}
the $\Psi(z)$ function must obeys
\begin{equation}
\label{schrodingerequation}
 -\ddot{\Psi}(z)+U(z)\Psi(z)=m^{2}\Psi(z),
\end{equation}
where
\begin{eqnarray}
U(z) & = & \frac{\ddot{\sigma}}{\sigma}+\frac{1}{2}\frac{\dot{\sigma}}{\sigma}\frac{\dot{\beta}}{\beta}-\frac{3}{16}\left(\frac{\dot{\beta}}{\beta}\right)^{2}+\frac{1}{4}\frac{\ddot{\beta}}{\beta}+\frac{l^{2}}{\beta}.
\end{eqnarray}
Equation $(\ref{schrodingerequation})$ is a time-independent Schr\"{o}dinger-like equation. We can study the localization of the scalar field by analyzing the behavior of the
potential around a potential well. Returning to $\rho$ coordinate the potential can be written as
\begin{eqnarray}
 U(\rho,k,l) & = & \sigma\Big(\frac{\sigma''}{\sigma}+\frac{1}{2}\left(\frac{\sigma'}{\sigma}\right)^{2}+\frac{\sigma'}{\sigma}\left(\frac{5}{8}\frac{\beta'}{\beta}\right)-\frac{3}{16}\left(\frac{\beta'}{\beta}\right)^{2}+\frac{1}{4}\frac{\beta''}{\beta}\Big)+\frac{l^{2}}{\beta}\\
     & = & k^{2}e^{-(k\rho-\tanh{k\rho})}\Big(\frac{3}{2}\tanh^{2}{k\rho}-\frac{9}{4}\text{sech}^{2}{k\rho}\tanh{k\rho}-\frac{1}{4}\frac{\text{sech}^{4}{k\rho}}{\tanh{k\rho}}-\text{sech}^{2}{k\rho}\Big)\nonumber\\
     & + & (kl)^{2}\frac{1}{\tanh^{2}{k\rho}}.
\end{eqnarray}
The Schr\"{o}dinger potential is plotted in fig. (\ref{cigarpotential}) for $l=0$. It is worth mention that there is a potential well and a barrier around the origin where the 3-brane is. Therefore, there are massive modes trapped to the brane.


\section{Conclusions and Perspectives}

\indent \indent In this work we have found an interior and exterior gravitational solution for a string-like brane in six dimensions. The metric was build from a warped product between a 3-brane and a two dimensional manifold called
Hamilton cigar soliton. Since the cigar soliton is a parameter dependent manifold with axial symmetry, that solution enable us to study the effects of variations of string-like brane configurations has on the physical quantities of the braneworld scenario, as the gravitational field, the relationship between the masses scales and so on, without break the own string-like axial symmetry.
Further, since that solution is valid for both interior and exterior region to the string-like brane, it provides an improvement on the analysis of the behavior of the fields inside and near the brane.

The bulk possesses a cosmological constant depending on the cigar parameter, also called evolution parameter. Thus, the bulk converges asymptotically to a constant curvature manifold which scalar curvature depends on the
evolution parameter. Besides, the stress-energy-momentum tensor also depends on the evolution parameter, satisfying the weak, strong and dominant energy condition for any value of $k$.

Another physical consequence is the parametrization of the relationship between the mass scales of bulk and brane. In order to take $M_{4} \gg M_{6}$, $k$ must be taken quite small. However, allowing $k$ to vary keeping $M_{6}$ fixed, the brane mass scale would vary which leads to both particle physics and cosmological consequences.

The differences between the string-cigar solution and the GS model arise due the former be a thick brane solution while the latter is a thin solution. As a result, the relationships between the string tensions,
that define the Tolman mass and the angular deficit, depend on the bulk cosmological constant and on the width of the core.

Due the resemblance to the GMS solution, we argue that the string-cigar should be a solution of some vortex model, for instance, a non-Abelian EMH model. Another possibility is this geometry be generated by a non-minimal
coupling between fields in a modified gravity theory, as currently done in cosmology. The deduction of this geometry from a matter lagrangian should be a future step in order to complete the model.

We have also studied the gravitational perturbations on that scenario. Since the volume of the transverse manifold is finite, the massless mode is trapped to the 3-brane for any value of the evolution parameter. In the interior and exterior regions but close to the brane, as higher the value of the evolution parameter higher the amplitude of the massless mode. For the Kaluza-Klein modes, the eigenfunction converge to the usual vacuum solution of GS model written in terms of the Bessel functions and then it is trapped to a brane up to a cutoff. The KK spectrum also depends on the evolution parameter since the KK mass depends on the cosmological constant. Nevertheless, the KK spectrum shift for a same amount since it depends only on the cosmological constant. The localization of the massive modes can also be inferred from shape of the analogous Schr\"{o}dinger-like potential. Indeed, there is a potential well and a potential barrier around the brane. Hence there are massive modes trapped to the brane
due the potential barrier. As a perspective, the spectrum of mass can be achieved numerically from this potential by means of the resonances method.

This article opens new perspectives to develop. We showed that steady geometric flows on the transverse manifold yields a variation of the relation between the Bulk and brane masses. We argue that,
whether this scenario be created by an Einstein-Maxwell-Higgs vortex, this mass flow can be linked with the Higgs and vector masses, since these are the mass content of the vortex. Therefore, for the
transverse space evolving under a general Ricci flow, it would be interesting investigate what physical quantities of the brane would be changing. For other fields living in that geometry, we could study the
effects that the evolution parameter has on their properties, as well the behavior of these fields on the region inside and near the string-like brane. Indeed, we could find for fermions, for instance, some
geometrical Yukawa potential or some different fermion generations. Since the brane has parameter dependent tensions, another possibility would be study the cosmological consequences of the geometric flow
has on the brane. Besides, the string-cigar extends the cigar-like universe due the presence of the cosmological constant. Then, a useful study could be deduce
the string-cigar from a K\"{a}hler potential, as done for the cigar-universe. Another interesting feature to be addressed is the gravitational modes behavior of sources non-minimally coupled to gravity, as
in the braiding models \cite{Deffayet:2010qz}, where the source interferes with the perturbation.


The authors would like to thank Conselho Nacional de Desenvolvimento Cient\'{\i}fico e
Tecnol\'{o}gico (CNPq) and  Coordena\c{c}\~{a}o de Aperfei\c{c}oamento de Pessoal de N\'{i}vel Superior (CAPES) for financial support.


\end{document}